# Approach to Visual Attractiveness of Event Space Through Data-Driven Environment and Spatial Perception

Aliffi Majiid[1][0009-0000-9287-3631], Riaz-Ul-Haque Mian[1][0000-0001-6550-5753],
Kouki Kurohara[1][0009-0008-5301-8163] and Yen-Khang Nguyen-Tran[1][0000-0002-0535-0179]

[1] Interdisciplinary Faculty of Science and Technology, Shimane University
`ifinmajid11@gmail.com`

**Abstract.** Revitalizing Japan's remote areas has become a crucial task, and Matsue City exemplifies this effort in its temporary event spaces, created through collective efforts to foster urban vibrancy and bring together residents and visitors. This research examines the relationship between data-driven insights using generative AI and visual attractiveness by evaluating temporary events in Matsue City, particularly considering the cognitive-cultural differences in processing visual information of the participants. The first phase employs semantic keyword extraction from interviews, categorizing responses into physical elements, activities, and atmosphere. The second phase analyzes spatial perception through three categories: layout hierarchy, product visibility, and visual attention. The correlation indicates that successful event design requires a balance between spatial efficiency and diverse needs, with a spatial organization that optimizes visitor flow and visibility strategies considering cultural and demographic diversity. These findings contribute to understanding the urban quality of temporary event spaces and offer a replicable framework for enhancing the visual appeal of events in remote areas throughout Japan.

# 1 Introduction

## 1.1 Background and Purpose

Revitalizing Japan's remote areas has become a crucial task [1], and Matsue City exemplifies this effort in its temporary event spaces, created through collective efforts to foster urban vibrancy and bring together residents and visitors. These events, where 70% of visitors expressed positive feedback, have continued giving way to other small communities' events promoting the city's potential and local brand among residents and tourists [2, 3]. Featuring organic foods, local products, family-oriented activities, etc. the diversity of stalls creates varied visual attractiveness to enrich participant's experience [4]. However, there are still concerns that temporary event spaces lack attraction and struggle to encourage participants to revisit in the long term due to inadequate engagement, often caused by a mismatch between event organizers' intentions and visitors' expectations. It highlighted a cognitive-cultural gap, showing significant differences between unmarried visitors and families at events, such as cultural exploration, novelty, socializing, and family togetherness [5]. In addition, people familiar with Japanese culture tend to use holistic processing strategies and overseas visitors focus independently on prominent objects, especially in event spaces where multiple visual elements compete for attention [6]. This situation intrigues the questions about how the spatial settings affect the participants' perception in the temporary event spaces, leading to defining the event's visual attractiveness in participants' interpretation from different cultures and demographics. Besides, in recent research, generative AI tools like ChatGPT-4 have demonstrated how natural language understanding can assist in analyzing natural voice conversation [7, 8]. The utilization of natural language processing (NLP) has also successfully assessed and decrypted user interpretations, providing advantages in accessing people's evaluations [9]. Furthermore, digitalization or AI has presented new challenges in adopting data-driven design paradigms to enhance user experiences [10]. By combining visual segmentation and spatial perception analysis with NLP, the findings will enhance the quality of temporary events and demonstrate design strategies in visual attractiveness at temporary event spaces for Matsue City and other remote areas, making a replicable framework for community events for remote areas in Japan.

# 2 Methodology

The analysis is structured in two steps. First, we evaluate the event space from the viewpoint of visitors using semantic keywords extracted through interview from two perspectives with and without predefined classifications using generative AI and our proposed adaptive keyword analysis model. Second, we adapt the viewpoint of organizer to categorize spatial configuration from layout hierarchy, products visibility, and visual attention. Finally, by correlating these two viewpoints, we evaluate and propose guidelines for optimizing temporary events in remote areas.

The data collection started by conducting observation and video recordings of three selected events in Matsue (EV-1, EV-2, EV-3) from August to October 2024. At the same time, event stall layouts and decoration elements were also mapped, together with store owners' design intentions. These on-site recordings were then edited into videos for a post-event interview during November 2024 with three long-term residents and three foreign visitors, three of them having children. The interview data is used to evaluate the attractiveness of event space from the viewpoint of participants and the mapping data is to evaluate from the viewpoint of organizers. Regarding the case studies, the targeted event was chosen in Matsue City, known for its cultural and historical richness and lively community building through event organizations. EV-1 is a monthly market organized to promote local shops; it accommodates between 20 to 30 visitors each time. EV-2 is a bi-annual community event that intends to provide additional events for the community during off-seasons, it accommodates between 30-50 visitors each time. EV-3 is the city monthly farmer market aiming to promote environmental awareness of business around Matsue, with close to a hundred visitors monthly. Despite differences in terms of scale, number of visitors and location, the three events all share the same characteristics of promoting local products and with the intention of creating a temporary event to boost urban liveliness.

## 3      Data-Driven Insight into Environmental Dynamics

The methodology for semantic keyword extraction and analysis was developed to systematically extract, categorize, and analyze keywords from interview data related to all events, ensuring they align with predefined classification criteria. To maintain objectivity and prevent any *bias* from prior interactions, the analysis began with a fresh session, effectively resetting the context to avoid interference. This analysis is conducted in two phases. In both phases, the input data was organized in a structured Microsoft Excel format, including columns that detailed participant information with name, nationality, parental status, gender, and their qualitative responses for each event. This structured approach facilitated consistency and traceability throughout the analysis process. Providing background information about each event was essential to contextualize the data. Details about the event's purpose, activities, and target audience helped ensure that the keyword extraction remained relevant and grounded in the specific context of each community event.

In *phase-1*, keywords were categorized into three distinct classifications: *Activities* (actions performed or observed, like eating, walking, buying), *Physical Elements* (tangible objects or items mentioned, such as sweets or farm products), and *Atmosphere* (environmental or emotional impressions, like lively, fun, or natural). Clear definitions for each category were established to maintain semantic precision and ensure that the keywords accurately reflected the content and nuances of the interviews. Then, a keyword extraction matrix was employed, mapping the keywords into a six-column structure comprising the keyword and its definition for each of the three categories: *Physical Elements, Activities,* and *Atmosphere*.

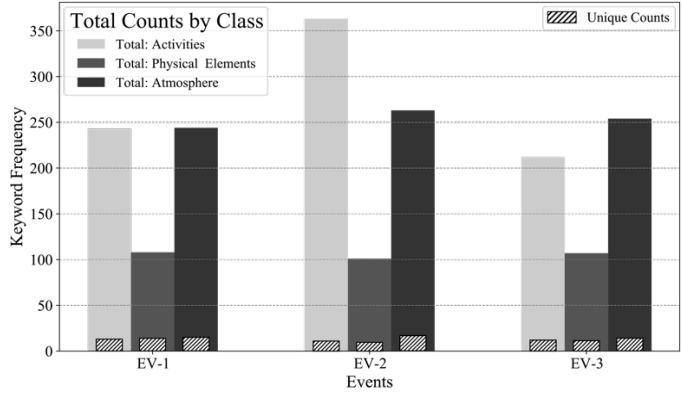

**Fig.1.** NLP-based Semantic Keyword Extraction using the GPT-4 Model API [8] *(Phase-1)*, illustrating unique and total counts across event categories

Definitions were crafted to highlight the key purpose and contextual relevance of each keyword, focusing on their core meanings and referencing the original context from the interviews. This step was crucial in preserving the semantic integrity and ensuring that each keyword was directly traceable back to the source text. Fig.1 summarizes the *phase-1* total (bar) and unique keyword frequencies (inner bar) of three categories across events.

In *phase-2*, the data was analyzed without predefined classification, and ChatGPT self-generated identification was adjusted to identify potential correlations with the *phase-1* outputs. Through structured methodology, interview content was analyzed using *keyword extraction*, *categorization*, and *thematic analysis*. In both phases, keywords were systematically extracted per person based on predefined criteria to ensure accuracy and relevance.

### 3.1 Proposed Algorithm for suitable Keyword Weighting in Text Analysis

In both phases, keyword frequency can be significantly biased due to human behavior, disproportionately impacting certain keywords. For instance, in Fig. 3, the keyword *"see"* in the activity category has a much larger influence compared to others. Similarly, Person-3 (the interviewer) exerts greater influence than others. To address these biases, keywords are identified using the GPT-4 API, categorized based on relevance, and their frequencies analyzed to determine occurrence. The proposed algorithm, illustrated in Fig. 2, calculates weights for each keyword within a defined range (*α_max* and *α_min*), streamlining classification and prioritization. $W_k$ is calculated using the formula Eq. 1, where $f_k$ represents the frequency of keyword $k$ and $\omega$ is the calculated weight factor.

$$W_k = \begin{cases} \alpha\_max, & \text{if } 1 + (f_k \times \omega) > \alpha\_max \\ \alpha\_min, & \text{if } 1 + (f_k \times \omega) < \alpha\_min \\ 1 + (f_k \times \omega), & \text{otherwise} \end{cases} \quad (1)$$

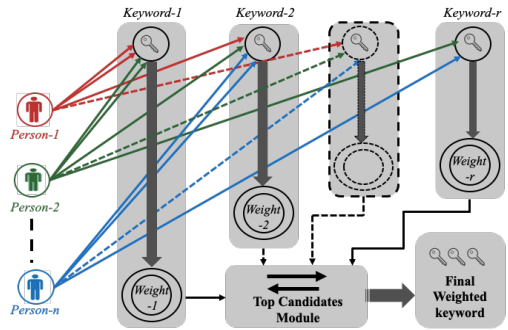

**Fig. 2.** Proposed adaptive keyword analysis model architecture

The proposed algorithm ensures fair weight distribution, assigning higher significance to frequently used keywords while maintaining proportionality. As shown in Fig. 3, the top seven keywords and each person's keyword frequency are adjusted based on the proposed algorithm illustrated in Fig. 2 with the weight factor ω=0.08.

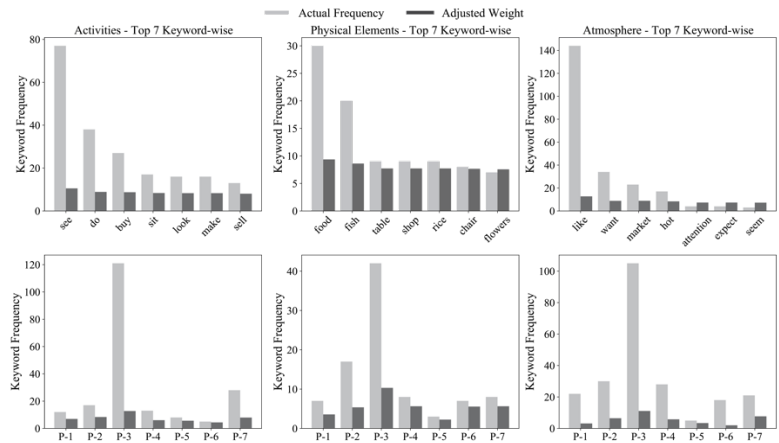

**Fig. 3.** Actual and adjusted Keyword frequencies for EV-1 using the proposed algorithm.

### 3.2 Analysis and Key Findings of Data-Driven Environment

After analyzing biases, this part evaluates feedback from all participants observing three events. At this stage, interview-related information is also considered to determine intentions and correlations with the *phase-1* keyword classification.

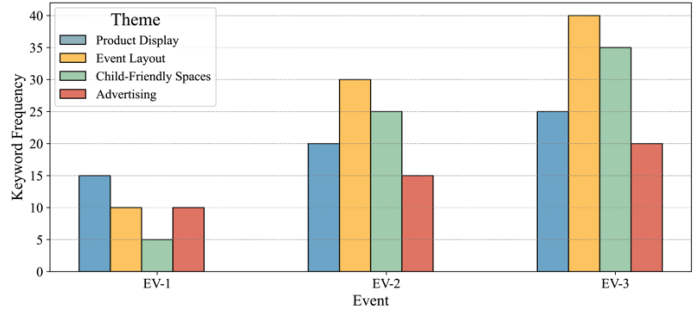

**Fig. 4.** Thematic suggestions for all events using the GPT-4 Model API and the proposed algorithm *(Phase-2)*.

Fig. 4 illustrates the thematic feedback distribution across all events, emphasizing improvements in product display, event layout, and child-friendly spaces, achieved through the combined use of the proposed adaptive keyword analysis model architecture and the GPT-4 API. These graphs illustrate the focus areas for enhancing event design and participant experience such as improving showcase presentation (Product Display), event arrangement or planning (Event Layout), kids' zones (Child-Friendly Spaces), and promotional activities (Advertising). These insights underscore the importance of design expertise and organizers in designing inclusive and functional event environments, supporting broader goals of community engagement, tradition preservation, and environmental awareness.

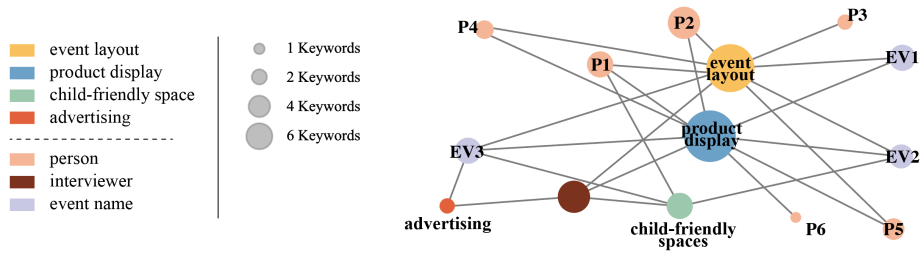

**Fig. 5.** Graphical representation of relationships among events, themes, and participants keyword weights are adjusted by the proposed model

Finally, Fig. 5 illustrates the outcome of the data-driven model, incorporating responses across three dimensions: event-wise themes and the impact of participants, adjusted for keyword weights. Circle sizes represent the relative impact of each entity (event, person, and theme). The proposed algorithm, supported by NLP techniques, ensures fair weight distribution and contextual relevance. Analysis of participant feedback highlights key areas for improvement, forming a comprehensive data-driven environment.

## 4 Spatial Perception impact on Visual Attractiveness

### 4.1 Categorization of Spatial perception

Building on users' perception insight from data-driven model for event space improvement, this study further examines spatial perception by classifying 42 stalls into three categories: Layout Hierarchy, Products Visibility, and Visual Attention, which correspond to Event Layout, Product Display, and Advertising, respectively. The first category identifies the positioning of shops advertisements, the second assesses the visibility of stalls from event entrances, the third investigates vendors' strategies in implementing advertising elements. The result will explain how vendors utilize those visual attractiveness strategies to enhance their shops by impacting visitors' perception.

Firstly, the layout hierarchy can be categorized into three types based on their display position (Fig. 6). LH1 features a display table within the stall periphery, while LH2 expands in front of the stall with additional advertisements, and LH3 adds more elements in the back space of the store, within the stall periphery.

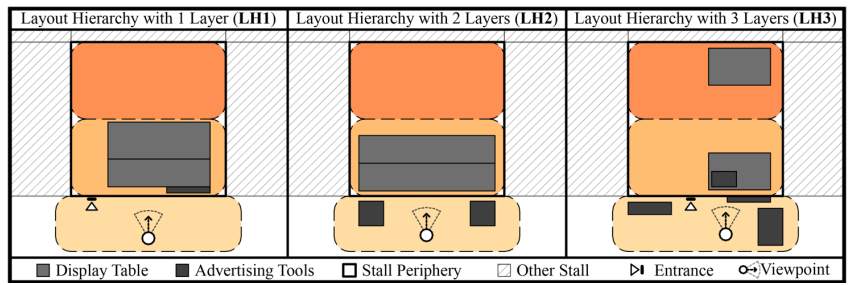

**Fig. 6.** Layout hierarchy classification

Then we look at how many of these layout hierarchies are present in EV-1, EV-2, and EV-3, which show a dominant of LH1 and LH2 with same quantity (Tab. 1).

**Tab.1.** Layout hierarchy in each event

| | Stall Sample |
|---|---|
| LH1 | EV1-02, EV1-06, EV1-07, EV1-09, EV1-10, EV2-01, EV2-09, EV2-10, EV2-11, EV3-03, EV3-09, EV3-10, EV3-12, EV3-15, EV3-16, EV3-20 |
| LH2 | EV1-01, EV1-03, EV1-05, EV2-03, EV2-05, EV2-07, EV3-01, EV3-04, EV3-05, EV3-06, EV3-07, EV3-08, EV3-11, EV3-13, EV3-14, EV3-21 |
| LH3 | EV1-04, EV1-08, EV2-02, EV2-04, EV2-06, EV2-08, EV3-02, EV3-17, EV3-18, EV3-19 |

Secondly, the level of visibility was ranging from the least to the most visible (PV1~PV3), corresponding to their visibility from the entrance of the event. In EV-1 layout, the stalls are arranged linearly with moderate visibility (PV2), while the indoor zone has PV1. In EV-2, the layout demonstrates a visibility distribution that highlights corner areas, with more stalls achieving high visibility thanks to their location on the periphery of the event area, while central stalls have PV2. The EV-3 features the most diverse visibility pattern due to its size, as visitors can access the location through multiple entrances. Central stalls maintain PV2, while peripheral stalls have PV3 due to strategic corner placement (Fig.7).

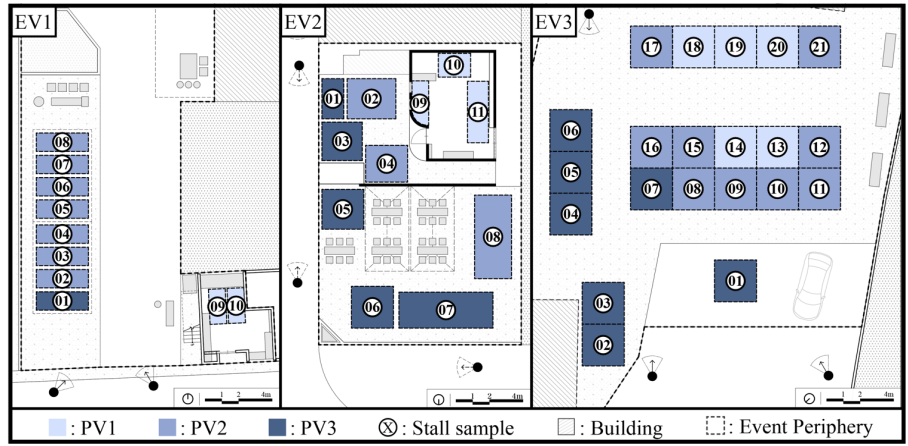

**Fig. 7.** Products visibility distribution in each event

Finally, the visual attention analysis reveals strategies for capturing visitors' impressions, involving position compared to eye level, banner information, size, and the color contrast on each store advertising elements. Fig. 8 shows that VA1 acquired the highest dominant with 17 stalls, suggesting a minimum placement of advertising elements, such as attaching them to tablecloths or placing the banner or product display on tables. The banners also vary in size with most of them displaying text and price as information. Next dominant is VA2 with 14 stalls, indicating shops using two banners at different positions in the shop front, the information also varies between text, photo and price. However, shops that use banners at various positions, bigger sizes and more contrast in color, such as VA3 are quite limited (7 stalls out of 42), as close as the quantity of VA0 (4 stalls) indicates the lack of display efforts to catch the attention of event visitors.

| Sample (42) | Position | | | Information | | | | | Banner size | Contrast color | Visual attention pattern | Visualization |
|---|---|---|---|---|---|---|---|---|---|---|---|---|
| | Below the table | Table to eye-level | Above the eye-level | Photo, text, price | Photo and text | Text only | Text and price | | | | | |
| | L1 | L2 | L3 | B1 | B2 | B3 | B4 | | BS | CC | VA | □ display  ■ advertising  ■ layout |
| EV2-09 | | | | | | | | | | | **VA0** No Implementation **(4)** | 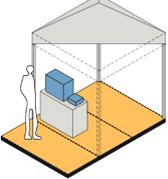 |
| EV1-09 | | | | | | | | | | ● | | |
| EV1-06 | ● | | | ● | | | | | | ● | | |
| EV3-10 | ● | | | ● | | | | | | ○ | ● | |
| EV3-07 | ● | | | | ● | | | | | ● | ● | **VA1** Single-level Implementation **(17)** | 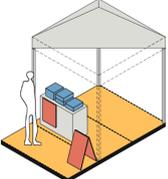 |
| EV1-01 | ● | | | | | ○ | | | | ● | | |
| EV2-04 | ● | | | | | | | | ● | ● | | |
| EV2-03 | ● | | | | | | | | ● | ○ | ● | |
| EV2-05 | ● | | | | | | | | ○ | ● | | |
| EV1-08 | ● | | | | | | | | ○ | ○ | | |
| EV3-03 | | ● | | ○ | | | | | | ○ | | |
| EV2-01 | | ● | | | | ● | | | | ● | | |
| EV2-10 | | ● | | | | ● | | | | ○ | ● | |
| EV1-07 | | ● | | | | | | | ● | ● | | |
| EV1-02 | | ● | | | | | | | ● | | ● | |
| EV1-10 | | ● | | | | | | | ○ | ○ | | |
| EV3-06 | | | ● | ● | | | | | | ● | | |
| EV1-05 | | | ● | | | | | | ○ | ○ | | |
| EV3-05 | ● | ● | | ● | | | | | | ● | ● | **VA2** Two-level Implementation **(14)** | 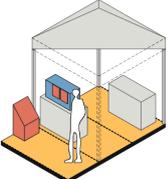 |
| EV3-11 | ● | ● | | ● | | | | | | ● | | |
| EV3-21 | ● | ● | | ● | | | | | | ○ | | |
| EV3-04 | ● | ● | | | ● | | | | | ● | ● | |
| EV3-17 | ● | ● | | | ● | | | | | ● | | |
| EV3-14 | ● | ● | | | | | | | ● | ● | | |
| EV3-13 | ● | ● | | | | | | | ● | ○ | ● | |
| EV1-03 | ● | ● | | | | | | | ● | ○ | | |
| EV3-02 | ● | | ● | ● | | | | | | ○ | | |
| EV3-18 | | ● | ● | | | ● | | | | ● | | |
| EV2-06 | ● | ● | ● | ● | | | | | | ● | ● | **VA3** Three-level Implementation **(7)** | 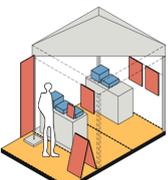 |
| EV1-04 | ● | ● | ● | ● | | | | | | ● | | |
| EV2-08 | ● | ● | ● | ● | | | | | | ○ | ● | |
| EV3-01 | ● | ● | ● | | ● | | | | | ● | ● | |
| EV3-19 | ● | ● | ● | | ● | | | | | ● | | |

**Fig. 8.** Visual Attention

### 4.2 Visual Attractiveness from Spatial Perception and Users Insights

The analysis of 42 stall samples across three event spaces in Matsue City reveals key insights into the impact of spatial perception on visual attractiveness. Vendors predominantly utilized single-layered (LH1) and double-layered (LH2) of layout hierarchy, which proved effective in organizing display and guiding visitors to identify key products. Corner and peripheral stalls benefited from their strategic placement along visitor pathways, consistently achieving higher visibility ratings. However, stalls in less visible locations (PV1, PV2) faced challenges in attracting attention and developed innovative solutions, such as elevated advertising tools placed outside tent peripheries and aligned with visitor sightlines. These less visible stalls frequently employed complex strategies (VA2, VA3), including vibrant color contrasts and engaging advertising tool to effectively redirect visitor focus.

By recognizing the potential of stalls across various scales, the design of successful event spaces requires careful consideration of architectural principles, user feedback, and diverse stakeholder interests. Particularly in remote areas with small populations and limited visitors, delivering outcomes that align with visitor expectations through improved event space quality can result in increased attendance while accommodating stakeholder priorities. Event organizers often pursue different goals, ranging from charity events for communities to profit-oriented, which directly influence their decisions regarding layout and vendor selection criteria. Through qualitative and quantitative analysis of visual attractiveness, the relationship between spatial perception and user insights becomes evident when designing event spaces. Layout hierarchy, product visibility, and visual attention significantly affect human perception in decision-making and stall awareness. Based on user insights, event layout frequently emerges as an area requiring improvement. This aligns with the challenges of implementing single-layered layout hierarchies (LH1), which struggle to engage visitors in recognizing key products, as they primarily focus on promoting items within the confines of individual tents. This underscores the necessity for organizers to strategically position anchor products at event layout corners and to optimize the proximity of grouped functions, aligning with behavioral patterns and demographic characteristics of the target audience. Furthermore, aside from child-friendly improvements emphasized in user insights, product display has been identified as the second most highlighted aspect. Despite limitations in visibility levels, stall owners inherently have equal opportunities to enhance their visual appeal through strategic approaches, such as utilizing a three vertical level advertising tool to achieve higher levels of awareness.

## 5    Conclusion

In conclusion, this study explores how visual attractiveness influences user experience in temporary event spaces in Matsue City, Japan, with a focus on revitalizing remote areas through community-driven events. By quantitatively extracting semantic keywords and qualitatively analyzing spatial perceptions, integrating ChatGPT's API with the proposed adaptive keyword analysis model, this research identifies three key success factors: layout hierarchy representing event layouts, product visibility as product displays, and visual attention toward advertising. The findings indicate that organizers should reorganize layouts by considering anchor points and proximity, ensuring that stalls with lower visibility still receive equal attention from visitors. Furthermore, vendors inherently have equal opportunities to optimize their marketing potential by emphasizing the importance of information completeness, readability, and appropriate sizes, complemented through color contrast with the surrounding environment. Finally, the results highlight actionable strategies to improve the quality of temporary events while offering a replicable framework for enhancing visual attractiveness and community engagement in Matsue City and similar remote areas in Japan.

The study limitations include a restricted data collection of a small interview sample size with imbalance gender and may not fully represent the diversity of temporary event spaces in other Japanese cities. Despite these constraints, the findings contribute valuable insights into the relationship between spatial configuration and visitor experience in temporary event spaces, particularly in Japanese suburban context. Future research could address these limitations by incorporating eye-tracking analysis and architectural software to regenerate the event spaces design based on interview feedback.